\documentstyle[aps,prb,preprint]{revtex}

\begin{document}
\draft
\title{A planar force-constant model for phonons in wurtzite GaN and AlN: Application to
hexagonal GaN/AlN supperlattices}
\author{Lingjun Wang and Guanghong Wei}
\address{Surface Physics Laboratory (National Key Lab), Fudan University,\\
Shanghai\\
200433, People's Republic of China}
\author{Jian Zi\cite{byline1}}
\address{CCAST (World Laboratory), P. O. Box 8731, Beijing 100080,\\
People's Republic of China and\\
Surface Physics Laboratory (National Key Lab), Fudan University, Shanghai\\
200433, People's Republic of China\cite{byline2}}
\date{\today}
\maketitle

\begin{abstract}
A planar force-constant model is developed for longitudinal phonons of
wurtzite GaN and AlN propagating along the [0001] direction. The proposed
model is then applied to the study of the phonon modes
in hexagonal (0001)-GaN/AlN supperlattices in longitudinal polarization. The
confinement of the superlattice phonon modes is discussed.
\end{abstract}


\narrowtext

\section{Introduction}

Among the III-V semiconductors the nitride-based semiconductors have
received both scientific and technological attention over the past years
owing to their excellent physical properties.\cite{str:92} The fascinating
mechanical properties, such as hardness, high melting temperature, high
thermal conductivity, and large bulk moduli make them useful for protective
coatings. In particular, the electronic properties, characterized by large
band gaps and relatively low dielectric constants, indicate tremendous
potential applications in optical devices working in the blue and
ultraviolet (UV) wavelengths. Recently, high-power blue and green light
emitting diodes (LEDs) have been fabricated by using InGaN-based
multiquantum-well structures and are commercially available. \cite{nak:95}
Blue and UV nitride semiconductor laser diodes (LDs) have also successfully
been demonstrated. \cite{nak:96}

Consequently there has been great interest in the fundamental properties,
the understanding of which can render the improvement of the materials and
device quality possible. Heterostructures based on GaN and Ga$_{1-x}$Al$_x$N
have become important candidates for optoelectronic materials. Great
attention has been paid to the nitride heterostructures and layered
structures. Short-period superlattices (SL's) have been successfully
fabricated by advanced growth techniques. \cite{asi:93} Phonon spectra of
these SL's were investigated by infrared (IR) spectroscopy. \cite{mac:96}

It has been known that GaN and AlN are crystallized in the stable wurtzite
structure and metastably in the cubic structure. Phonons in cubic GaN/AlN
systems have drawn interest in recent years. Theoretical studies on phonons
in cubic GaN/AlN SL's have been reported by several groups. \cite
{gri:94,gri:96,zi:96,wei:97} There have been, however, few studies of
phonons in hexagonal GaN/AlN SL's,\cite{wei:98a} where the constituents have
wurtzite structures.

In this paper we develop a planar force-constant model (PFCM) to describe
the longitudinal phonons in wurtzite GaN and AlN propagating along the
[0001] direction. We then apply the PFCM to study the phonons in hexagonal
(0001)-GaN/AlN SL's. The paper is organized as follows. The PFCM is
described in Section II. The calculated results for hexagonal (0001)-GaN/AlN
SL's are presented in Section III and conclusions are given in Section IV.

\section{Planar force-constant model}

In general, a three-dimension calculation is necessary in the discussion of
phonons because of the coupling between longitudinal and transverse
vibrations. For the vibrations of wurtzite compounds along the [0001]
direction, however, the longitudinal and transverse vibrations can be
decoupled. As a result, the longitudinal and transverse vibrations can be
treated separately.

The PFTM for wurtzite GaN or AlN is schematically displayed in Fig. \ref
{fig1}. The dynamical matrix elements are given by 
\begin{eqnarray}
D(q;\kappa ,\kappa ^{\prime }) &=&\frac 1{\sqrt{M_\kappa M_{\kappa ^{\prime
}}}}\sum_{l^{\prime }}\phi \left( l\kappa ;l^{\prime }\kappa ^{\prime
}\right) \times  \nonumber \\
&&\exp \left\{ iq\left[ U(l\kappa )-U(l^{\prime }\kappa ^{\prime })\right]
\right\} ,  \label{dy}
\end{eqnarray}
where $q$ is the wave-vector; $l$ is the index of a unit cell and $\kappa $
stands for the index of atoms in the unit cell; $M_\kappa $ is the mass of
the atom $\kappa $; $\phi \left( l\kappa ;l^{\prime }\kappa ^{\prime
}\right) $ is the planar force-constant between the atoms $(l\kappa )$ and $%
(l^{\prime }\kappa ^{\prime })$; $U(l\kappa )$ is the position of the atom $%
(l\kappa )$. The dynamical matrix subjects to the following secular equation

\begin{equation}
|D(q;\kappa \kappa ^{^{\prime }})-\delta _{\kappa \kappa ^{\prime }}\omega
^2(q)|=0.  \label{secular}
\end{equation}
The eigenvalues and eigenvectors of the vibrational modes can be obtained by
solving Eq. (\ref{secular}).

Four planar force-constants shown in Fig. \ref{fig1} are used for either GaN
or AlN. They are two cation-anion planar force-constants, one cation-cation
force-constant and one anion-anion force-constant. The on-site
force-constants $\phi (l\kappa ;l\kappa )$ can be obtained from the sum
rules 
\begin{equation}
\phi (l\kappa ;l\kappa )=-\sum\limits_{l^{^{\prime }}\kappa ^{^{\prime
}}(\neq l\kappa )}\phi (l\kappa ;l^{^{\prime }}\kappa ^{^{\prime }}).
\end{equation}

The values of the four planar force-constants are obtained by a fitting
procedure. The existing experimental data that are useful in fitting are the
A$_1$(LO) phonon frequency at the zone-center determined by Raman
spectroscopy. Therefore, the four planar force-constants are obtained by
fitting them to the three-dimensional calculations.\cite{wei:98b} The fitted
A$_1$(LO) phonon frequency, two B$_1$ phonon frequencies and the elastic
constant $C_{33}$ are given in Table \ref{tab1} together with the
corresponding experimental data and other theoretical results. It is obvious
that the fitted A$_1$(LO) phonon frequencies of both GaN and AlN are in good
agreement with the Raman experimental data. The two B$_1$ phonon frequencies
agree fairly well with other theoretical calculation. The fitted elastic
constants $C_{33}$ are also in good agreement with the experiment. The
resulting four planar force-constants are given in Table \ref{tab2}.

The calculated phonon dispersion based on the obtained planar
force-constants for wurtzite GaN and AlN for the longitudinal polarization
along the [0001] direction are shown in Fig. \ref{fig2}. It is found that
the four planar-force-constants can give very satisfactory phonon dispersion
of both GaN and AlN compared with other three-dimensional calculations.\cite
{wei:98b}

\section{Calculated results}

From Fig. \ref{fig2} some general features of phonon modes in hexagonal
GaN/AlN SL's can be postulated. If a SL mode is located inside GaN or AlN
continuum while GaN and AlN phonon continua do not overlap, vibrations of
this mode will sharply be confined to one of the constituent layers since
the other constituent cannot sustain this kind of vibrations. If a mode is
located in a region where GaN and AlN phonon continua overlap the mode will
be a resonant, quasi-confined one because the vibrations in another
constituent layers will be excited. If there is a SL mode with its frequency
lies neither in GaN continuum nor AlN continuum the vibrations of this mode
will be constrained in the interface region. It is a microscopic interface
mode.

With the PFCM introduced in the above section phonons in any hexagonal
(0001)-(GaN)$_m$/(AlN)$_n$ SL's can be studied, where $m$ and $n$ are the
number of bilayers\cite{bilayer} of GaN and AlN, respectively. For GaN/AlN
SL's, the planar force-constants are assumed to have the same values as
corresponding bulk except that across an interface the planar
force-constants are taken to be the average of those for GaN and AlN.

The calculated phonon dispersion and vibrational patterns is given in Fig. 
\ref{fig3} for a (0001)-(GaN)$_5$/(AlN)$_5$ SL. It is obvious that the
vibrations of modes 1-3 are sharply confined to AlN layers. They are
AlN-like modes originated from the A$_1$ branch of AlN. There is nearly no
dispersion for these modes due to the strong confinement of vibrations. Mode
4 is also an AlN-like mode, but it is from the optical B$_1$ branch of AlN.
The frequency of mode 5 is slightly above the A$_1$ branch of GaN. It is an
AlN-like mode, from the B$_1$ branch of AlN. Finite excitations in GaN
layers exist near the interfaces. Modes 6-9 are resonant, quasi-confined
GaN-like modes due to the fact that their frequencies are inside the overlap
region of the B$_1$ branch of AlN and A$_1$ branch of GaN. Vibrations in AlN
layers could be excited. Consequently, there exists finite dispersion for
these modes. Mode 10 is a GaN-like mode with vibrations confined to GaN
layers. Modes 11-13 are AlN-like modes and they are from the acoustic B$_1$
branch of AlN. The remaining modes are folded acoustic modes. The minigaps
in the zone-center and zone-boundary are due to the difference of the
elastic properties of GaN and AlN.

No interface modes are found in hexagonal GaN/AlN SL's. This may stem from
the fact that there are no {\it new bonds} across the interface.

For a confined mode the vibrations are sharply confined to one of
constituent layers. In consequence, the confined mode can be viewed as the
standing wave, described by an effective bulk wave-vector 
\begin{equation}
q_j=j\frac \pi {(n+\delta )d_0},  \label{conf}
\end{equation}
where $n$ is the nominal number of bilayers, $j$ is the order of a confined
mode, $d_0$ is the nearest-neighbor cation-cation or anion-anion distance
and $\delta $ is a parameter describing the degree of penetration of the
vibrations of a confined mode into the adjacent constituent. The frequency
of the confined mode can be derived simply from the bulk dispersion 
\begin{equation}
\omega _{\text{SL}}=\omega (q_j)
\end{equation}
instead solving the secular equation about the dynamical matrix if the
parameter $\delta $ is properly chosen. The above discussions are only valid
for the confined modes not for the resonant, quasi-confined modes or the
extended modes.

To test the above idea we give in Fig. \ref{fig4} the comparison of the bulk
dispersion and the frequencies of the confined modes of a (GaN)$_{10}$/(AlN)$%
_{10}$ SL mapped according to Eq. (\ref{conf}). A good correspondence is
established between the confined modes and the bulk dispersions for both GaN
and AlN when $\delta $ is chosen to be 1 for all branches. This means that
the vibrations of GaN or AlN confined modes extend about one bilayer into
the adjacent constituent. For those resonant, quasi-confined modes
significant vibrations exist in the other constituent layers. Their
frequencies are somewhat dependent on the thickness of the constituents.
Therefore standing wave picture is no longer valid. This can be clearly seen
from Fig. \ref{fig4}.

\section{\bf Conclusions}

In the present work a PFCM is developed for wurtzite GaN and AlN for the
longitudinal phonons propagating along the [0001] direction. We apply the
PFCM to the study of the phonons in hexagonal (0001)-GaN/AlN SL's. There are
confined modes and resonant, quasi-confined modes depending on their
frequencies. No interface modes are found in this system. The vibrational
confinement is investigated. For confined modes either GaN-like or AlN-like
their vibrations can be viewed as standing waves described by an effective
wave-vector. Their frequencies can be obtained directly from the bulk
dispersion.

\acknowledgements This work is supported from the NNSF of China under
Contract No. 69625609. Partial support from Qi-ming Star Project of
Shanghai, China is acknowledged.

\begin{table}[tbp]
\caption{Comparison of the fitted phonon frequencies at the zone-center and
elastic constants of wurtzite GaN and AlN with other experimental and
theoretical results. The phonon frequencies are in units of cm$^{-1}$ and
elastic constants are in units of GPa.}
\label{tab1}
\begin{tabular}{llllll}
&  & $\omega$(A$_1$) & $\omega$(B$_1^{(1)}$) & $\omega$(B$_1^{(2)}$) & $%
C_{33}$ \\ 
\tableline GaN & This work & 739 & 690 & 330 & 399 \\ 
& Others & 735\tablenotemark[1], 736\tablenotemark[2] & 677\tablenotemark[3]%
, 697\tablenotemark[4] & 330\tablenotemark[3], 335\tablenotemark[4] & 398%
\tablenotemark[5] \\ 
AlN & This work & 893 & 718 & 542 & 390 \\ 
& Others & 893\tablenotemark[6], 891\tablenotemark[7] & 717\tablenotemark[3]%
, 703\tablenotemark[4] & 553\tablenotemark[3], 534\tablenotemark[4] & 389%
\tablenotemark[8]
\end{tabular}
\tablenotemark[1]{Raman measurement, from Ref. \onlinecite{azu:95}.}\newline
\tablenotemark[2]{Raman measurement, from Ref. \onlinecite{koz:94}.}\newline
\tablenotemark[3]{First-principles full-potential LMTO calculations, from
Ref. \onlinecite{gor:95}.}\newline
\tablenotemark[4]{First-principles pseudopotential calculations, from Ref. %
\onlinecite{miw:93}.}\newline
\tablenotemark[5]{Measurement, from Ref. \onlinecite{pol:96}.}\newline
\tablenotemark[6]{Raman measurement, from Ref. \onlinecite{fil:96}.}\newline
\tablenotemark[7]{Raman measurement, from Ref. \onlinecite{hay:91}.}\newline
\tablenotemark[8]{Measurement, from Ref. \onlinecite{mcn:93}.}
\end{table}

\begin{table}[tbp]
\caption{Values of the planar force-constants obtained by a fitting
procedure for wurtzite GaN and AlN in units of Nm$^{-1}$.}
\label{tab2}
\begin{tabular}{ccc}
& GaN & AlN \\ 
\tableline $k_1$ & 22.45 & 23.69 \\ 
$k_2$ & 14.96 & 19.55 \\ 
$k_{3c}$ & 2.30 & 1.07 \\ 
$k_{3a}$ & 0.36 & -0.34
\end{tabular}
\end{table}

\begin{figure}[tbp]
\caption{Schematic of the PFCM for wurtzite GaN or AlN along the [0001]
direction.}
\label{fig1}
\end{figure}

\begin{figure}[tbp]
\caption{Calculated bulk dispersion of wurtzite GaN (dotted lines) and AlN
(solid lines) for the longitudianl phonons propagating along the [0001]
direction based on the PFCM. The phonon symmetry of GaN at the zone-center
is given.}
\label{fig2}
\end{figure}

\begin{figure}[tbp]
\caption{(a) Calculated phonon dispersion of a (0001)-oriented (GaN)$_5$/(AlN)%
$_5$ SL for phonons propagating along the [0001] direction. Here, $d$ is the
period of the SL. (b) The corresponding atomic displacement patterns at the
zone-center. The modes are numbered from the top in order of decreasing
frequency. For clarity, the displacements are displayed perpendicular to the
[0001] direction although they are actually pararell to the [0001] direction.
The dotted line represents interficial N atomic plan.}
\label{fig3}
\end{figure}

\begin{figure}[tbp]
\caption{Comparison of the bulk phonon dispersion of wurtzite GaN and
AlN along the [0001] direction and the confined-mode
frequencies of a (0001)-(GaN)$_{10}$/(AlN)$_{10}$ SL mapped according to Eq. (\ref
{conf}).}
\label{fig4}
\end{figure}


\begin{references}
\bibitem[*]{byline1}  To whom all correspondence should be addressed. Email:
jzi@fudan.edu.cn.

\bibitem[\dag]{byline2}  Mailing address.R. F. Davis, Proc. IEEE {\bf 79},
702 (1991).

\bibitem{str:92}  S. Strite and H. Morkoc, J. Vac. Sci. Technol. B {\bf 10},
1237 (1992).

\bibitem{nak:95}  S. Nakamura, M. Senoh, N. Iwasa, and S. Nagahama, Jpn. J.
App. Phys., Part 2 {\bf 34}, L797 (1995); S. Nakamura, M. Senoh, N. Iwasa,
S. Nagahama, T. Yamada, and T. Mukai, {\it ibid}. {\bf 34}, L1332 (1995).

\bibitem{nak:96}  S. Nakamura, M. Senoh, S. Nagahama, N. Iwasa, T. Yamada,
T. Matsushita, H. Kiyoku, and Y. Sugimoto, Jpn. J. Appl. Phys., Part 2 {\bf %
35}, L74 (1996); {\it ibid}., L217 (1996); K. Itaya, M. Onomura, J. Nishio,
L. Sugiura, S. Saito, M. Suzuki, J. Rennie, S. Nunoue, M. Yamamoto, H.
Fujimoto, Y. Kokubun, Y. Ohba, G. Hatakoshi, and M. Ishikawa, {\it ibid}.,
L1315 (1996); I. Akasaki, S. Sota, H. Sakai, T. Tanaka, M. Koike, and H.
Amano, Electron. Lett. {\bf 32}, 1105 (1996).

\bibitem{asi:93}  M. Asif Khan, J. N. Kuznia, D. T. Olson, T. George, and W.
T. Pike, Appl. Phys. Lett. {\bf 63}, 3470 (1993).

\bibitem{mac:96}  M. F. MacMillan, R. P. Devaty, W. J. Choyke, M. Asif Khan,
and J. N. Kuznia, J. Appl. Phys. {\bf 80}, 2372 (1996).

\bibitem{gri:94}  H. Grille and F. Bechstedt, Superlatt. \& Microstruct. 
{\bf 16}, 29 (1994).

\bibitem{gri:96}  H. Grill and F. Bechstedt, J. Raman Spectr. {\bf 27}, 201
(1996).

\bibitem{zi:96}  J. Zi, G. Wei, K. Zhang, and X. Xie, J. Phys.: Condens.
Matt. {\bf 8}, 6329 (1996).

\bibitem{wei:97}  G. Wei, J. Zi, K. Zhang, and X. Xie, J. Appl. Phys. {\bf 82%
}, 622 (1997).

\bibitem{wei:98a}  G. Wei, J. Zi, K. Zhang, and X. Xie, unpublished.

\bibitem{wei:98b}  G. Wei, J. Zi, K. Zhang, and X. Xie, Acta Phys. Sin.
(Overseas Ed.) {\bf 7}, 841 (1998).

\bibitem{bilayer}  A cation monolayer and an anion monolayer are counted as
a bilayer.

\bibitem{azu:95}  T. Azuhata, T. Sota, S. Suzuki, and S. Nakamura, J. Phys.:
Condens. Matt. {\bf 7}, L129 (1995).

\bibitem{koz:94}  T. Kozawa, T. Kachi, H. Kano, Y. Taga, and M. Hashimoto,
J. Appl. Phys. {\bf 75}, 1098 (1994).

\bibitem{gor:95}  I. Gorczyca, N. E. Christensen, E. L. Peltzer y Blanca,
and C. O. Rodriguez, Phys. Rev. B {\bf 51}, 11936 (1995).

\bibitem{miw:93}  K. Miwa and A. Fukumoto, Phys. Rev. B {\bf 48}, 7897
(1993).

\bibitem{pol:96}  A. Polian, M. Grimsditch, and I. Grzegory, J. Appl. Phys. 
{\bf 79}, 3343 (1996).

\bibitem{fil:96}  L. Filippidis, H. Siegle, A. Hoffmann, C. Thomsen, K.
Karch, and F. Bechstedt, Phys. Stat. Sol. (b) {\bf 198}, 621 (1996).

\bibitem{hay:91}  K. Hayashi, K. Itoh, N. Sawaki, and I. Akasaki, Solid
State Commun. {\bf 77}, 115 (1991).

\bibitem{mcn:93}  L. E. McNeil, M. Grimsditch, and R. H. French, J. Am.
Ceram. Soc. {\bf 76}, 1132 (1993).
\end{references}
\end{document}